\documentclass[10pt,conference]{IEEEtran}
\usepackage{amsmath,amsfonts}
\usepackage{algorithmic,bm}
\usepackage{array}
\usepackage{textcomp}
\usepackage{stfloats}
\usepackage{url}
\usepackage{verbatim}
\usepackage{graphicx}
\def\BibTeX{{\rm B\kern-.05em{\sc i\kern-.025em b}\kern-.08em
    T\kern-.1667em\lower.7ex\hbox{E}\kern-.125emX}}

\usepackage{lineno}

\usepackage{setspace}

\usepackage{balance}
\usepackage{commath}
\usepackage{cancel}
\usepackage{algorithm}
\usepackage{cuted} 
\usepackage{epsfig,epstopdf,graphicx,subfigure,psfrag,amsmath,cases}
\usepackage{latexsym,amssymb,amsmath,epsfig,subfigure,algorithm,amsthm}
\usepackage{nopageno}
\usepackage{algorithmic}
\usepackage{color}
\usepackage{url}
\usepackage{scrtime}
\usepackage{bm}
\usepackage{graphicx}
\usepackage{bbding}
\usepackage{multicol}
\usepackage{graphicx}
\usepackage{amsthm}
\usepackage{cite}
\usepackage{array}
\usepackage[utf8]{inputenc}
\usepackage[english]{babel}
 \usepackage{setspace}
 \usepackage{lipsum}
\addto\captionsenglish{\renewcommand{\figurename}{Fig.}}
{}
\newtheorem{theorem}{Theorem}

\usepackage{geometry}
\begin{document}
  \title{Sum-Rate Maximization for Pinching Antenna-assisted NOMA Systems with Multiple Dielectric Waveguides}\thispagestyle{empty}
 \author{
\IEEEauthorblockN{Shaokang Hu,~\IEEEmembership{Member,~IEEE}, 
Ruotong Zhao,~\IEEEmembership{Student Member,~IEEE}, 
Yihuan Liao,~\IEEEmembership{Member,~IEEE}, \\
Derrick Wing Kwan Ng,~\IEEEmembership{Fellow,~IEEE}, 
and Jinhong Yuan,~\IEEEmembership{Fellow,~IEEE}}

\IEEEauthorblockA{School of Electrical Engineering and Telecommunications, 
The University of New South Wales, Sydney, NSW 2052, Australia}  
\thanks{D. W. K. Ng and S. Hu are supported by the Australian Research Council (ARC)'s Discovery Projects DP240101019 and DP230100603, respectively. \par
J. Yuan is partially supported by the ARC's Discovery Projects DP220103596 and by the  ARC's Linkage Project LP200301482.}
}

\IEEEoverridecommandlockouts
\maketitle
\renewcommand{\IEEEaftertitletext}{\vspace{-0mm}}  
\begin{abstract}
This paper investigates the resource allocation design for a pinching antenna (PA)-assisted multiuser multiple-input single-output (MISO) non-orthogonal multiple access (NOMA) system featuring multiple dielectric waveguides. To enhance model accuracy, we propose a novel frequency-dependent power attenuation model for the dielectric waveguides in PA-assisted systems.
By jointly optimizing the precoding vector and the PA placement, we aim to maximize the system's sum-rate while accounting for the power attenuation across the dielectric waveguides. The design is formulated as a non-convex optimization problem. To effectively address the problem at hand, we introduce an alternating optimization-based algorithm to obtain a suboptimal solution in polynomial time. Our results demonstrate that the proposed PA-assisted system not only significantly outperforms the conventional system but also surpasses a naive PA-assisted system that disregards power attenuation. The performance gain compared to  the naive PA-assisted system becomes more pronounced at high carrier frequencies, emphasizing the importance of considering power attenuation in system design.
\end{abstract}


\section{Introduction}
The upcoming sixth-generation (6G) communication network is anticipated to deliver ultra-fast data transmission and seamless connectivity \cite{zhang20196g,10633213}. To fulfill the stringent quality of service (QoS) requirements of 6G, various cutting-edge technologies, such as intelligent reflecting surfaces (IRSs) \cite{zhang2021active,9838902,9505267,10104572}, fluid antenna systems \cite{wong2020fluid}, and movable antennas \cite{zhu2023movable,11048972}, have demonstrated their effectiveness in enhancing system capacity by proactively modifying channel environments. Indeed, these innovations enable channel characteristics to be dynamically tailored to cope with evolving communication requirements and to mitigate interference from other users or systems, thereby significantly improving system performance, increasing data rates, and enhancing reliability.
Specifically, movable antenna and fluid antenna systems constitute pioneering solutions that offer spatial diversity by implementing subtle adjustments to their physical positions \cite{zhu2023movable} or electromagnetic properties \cite{wong2020fluid}. However, the range of motion of these systems is typically limited to only a few wavelengths, resulting in only a marginal impact on mitigating large-scale path losses. Indeed, when the line-of-sight (LoS) link is blocked, these minor antenna position adjustments are often inadequate to restore it, leading to a substantial degradation in communication performance.
Moreover, IRSs can reflect signals to circumvent obstacles and establish a virtual LoS link by being deployed between transceivers. While typical IRS-assisted systems offer enhanced flexibility, they inherently suffer from “double path loss” signal attenuation \cite{zhang2021active}, weakening the overall signal strength. Thus, to further enhance the system data rate, establishing a low-cost, highly adaptable LoS link is imperative.

Recently, pinching antennas (PAs) were first introduced by DOCOMO in 2022 \cite{suzuki2022pinching}, presenting a promising solution to establish strong LoS communications. The fundamental concept behind PAs is to leverage dielectric waveguides in conjunction with antennas that employ a distinct dielectric material to effectively pinch the waveguides \cite{ding2024flexible,zhao2025resource}. In this configuration, radio waves propagating along the dielectric waveguides can be channeled into the surroundings of the PAs, which then radiate the signal to establish a designated communication zone. One of the distinctive features of PAs is their ability to slide along the dielectric waveguides, allowing them to radiate radio waves from any point on the waveguides. Moreover, once the PAs are removed, the corresponding radiation can be terminated \cite{suzuki2022pinching}. Compared to conventional fixed-location antenna systems, e.g., \cite{9738442,7277111}, PAs offer significant benefits, such as flexible deployment at chosen points along the waveguides, enabling efficient on-demand LoS link establishment to desired users. 

Several studies in the literature have demonstrated that PAs significantly enhance system performance across various communication systems. For instance, the authors of \cite{tegos2024minimum} maximized the minimum achievable data rate for uplink PA systems, confirming their superiority over fixed antenna approaches. Since PAs activated on a given waveguide must transmit an identical signal, traditional spatial multiplexing techniques become inadequate when the number of users exceeds the number of available waveguides \cite{ding2024flexible}. To address this, non-orthogonal multiple access (NOMA) is employed, allowing multiple users to share the same beam, thereby accommodating more users than the hardware typically permits while providing increased degrees of freedom (DoF) \cite{9804220}. 
Furthermore, NOMA is implemented by ordering users based on their channel qualities, with more power allocated to those experiencing weaker channel conditions. Notably, in conventional antenna-based NOMA systems, channel qualities are fixed and cannot be adjusted, inherently limiting both capacity and resource allocation fairness. 
In contrast, PA systems allow for the customization of effective channel qualities by strategically repositioning antennas. This flexibility facilitates better optimization of power allocation and user ordering, resulting in improved overall system capacity.
Inspired by the numerous advantages, the authors of \cite{wang2024antenna} investigated a NOMA-assisted downlink PA system, where multiple PAs along a dielectric waveguide serve multiple users via NOMA. However, both studies  \cite{tegos2024minimum} and \cite{wang2024antenna} are limited to single-waveguide configurations without extending the setup to a more general multi-waveguide PA system. 
To further generalize the PAs system model, the authors of \cite{bereyhi2025downlink} explored the deployment of PA systems with multiple waveguides to maximize the achievable weighted sum-rate. 
Note that multiple dielectric waveguides not only provide higher DoF, but also enable transmitting different information streams simultaneously. While these systems offer advancements, they also present new challenges.
For instance, existing works \cite{tegos2024minimum, bereyhi2025downlink} have idealistically ignored the power attenuation along the waveguide, which can reduce signal strength, potentially degrading overall system performance \cite{elements-electromagnetics}. 
In reality, power attenuation typically increases as carrier frequency rises \cite{suzuki2022pinching}, which is particularly significant in PA-assisted systems, as they are specifically designed to operate at high carrier frequencies, such as those in millimeter-wave bands \cite{suzuki2022pinching}. Thus, developing an accurate model for power attenuation in waveguides is essential.

Motivated by these observations, this paper investigates a PA-assisted multiuser multiple-input single-output (MISO) NOMA wireless system utilizing multiple dielectric waveguides. Our contributions are as follows:
i) We develop a model for power attenuation along dielectric waveguides in PA-assisted systems, representing a significant advancement toward a more realistic and practical system design for PA-assisted wireless networks;
ii) We formulate a resource allocation problem to maximize the system sum-rate by jointly optimizing the precoding vector and the location of the PAs, which enhances system performance;
iii) The formulated problem is non-convex, posing significant computational challenges. To address this, we propose a computationally efficient alternating optimization (AO)-based algorithm that yields an effective suboptimal solution;
iv)  Our results demonstrate substantial performance improvements with PAs over conventional antennas and the native PA system, which neglects power attenuation along dielectric waveguides. Moreover, the results highlight the critical importance of accounting for power attenuation, particularly at high carrier frequencies.

\emph{Notations}: Scalars, vectors, and matrices are denoted by \( x \), \( \mathbf{x} \), and \( \mathbf{X} \), respectively. 
\( \mathbb{R}^{N \times M} \) and \( \mathbb{C}^{N \times M} \) represent real and complex \( N \times M \) matrices, respectively. 
\( \mathbb{R}^{+} \) denotes positive real numbers, and \( \mathbb{H}^N \) is the set of \( N \times N \) Hermitian matrices. 
The modulus of a complex scalar is \( |\cdot| \) and its conjugate is \( x^* \). Euclidean norm of a vector is $\|\cdot\|$.
The transpose, conjugate transpose, expectation, rank, and trace of \( \mathbf{X} \) are \( \mathbf{X}^{\mathrm{T}} \), \( \mathbf{X}^{\mathrm{H}} \), \( \mathbb{E}\{\mathbf{X}\} \), \( \mathrm{Rank}(\mathbf{X}) \), and \( \mathrm{Tr}(\mathbf{X}) \), respectively. 
\( \mathbf{X}[q,i] \) is the element at row \( q \), column \( i \). 
\( \min\{a,b\} \) returns the smaller of \( a \) and \( b \). 
\( \mathbf{X} \succeq \mathbf{0} \) indicates that \( \mathbf{X} \) is positive semi-definite. 
$\mathcal{R}\{\cdot\}$ denotes the real part of a complex number. \( j \) is the imaginary unit. 
A circularly symmetric complex Gaussian (CSCG) random variable with mean \( \mu \) and variance \( \sigma^2 \) is denoted as \( \mathcal{CN}(\mu, \sigma^2) \), with \( \sim \) meaning “distributed as”.

\section{System Model}
\begin{figure}[t] 
  \centering
  \includegraphics[width=0.42\textwidth]{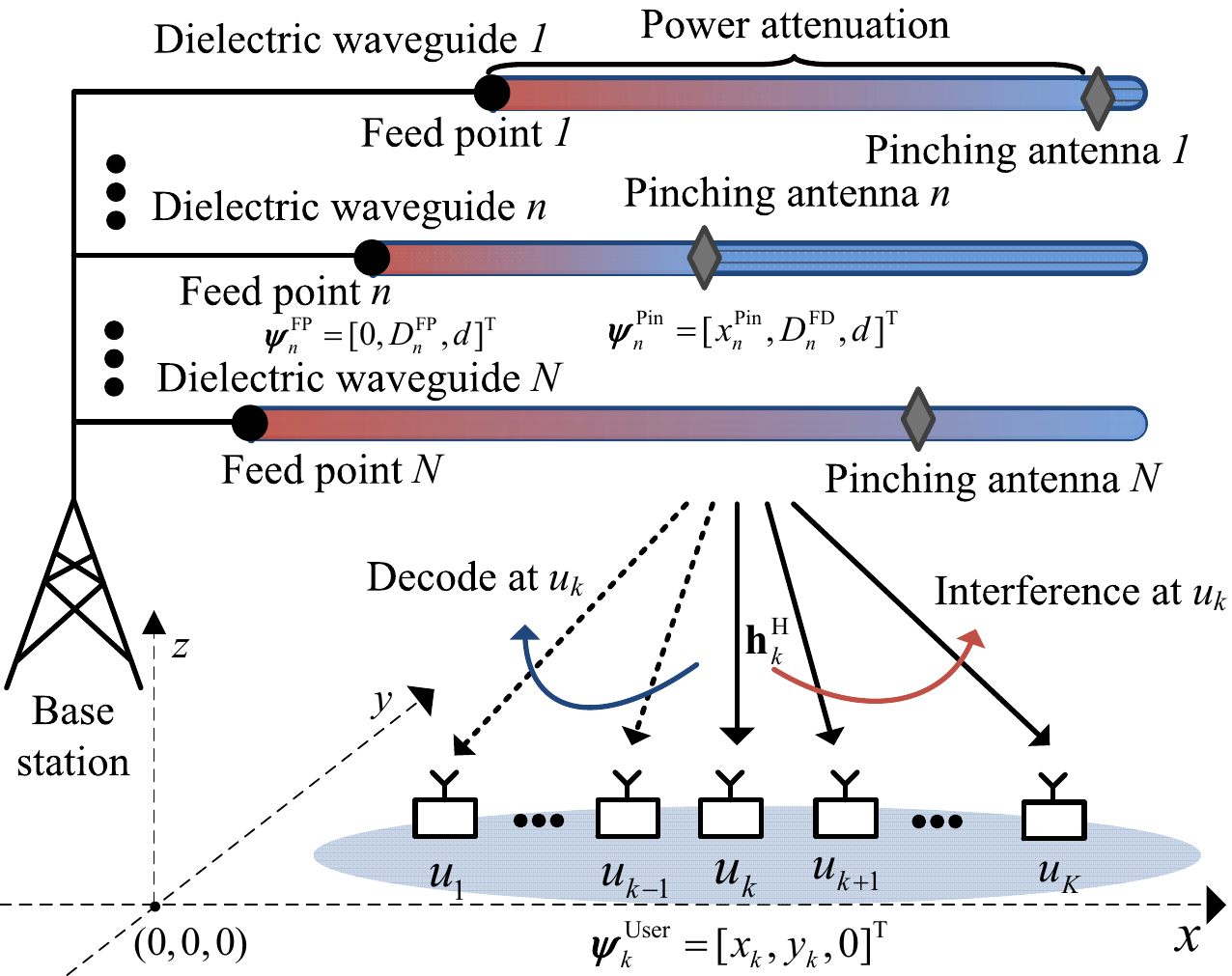}
  \caption{A downlink PAs-assisted communication system.}
  \label{fig:sys_mod}
\end{figure}
As illustrated in Fig.~\ref{fig:sys_mod}, we consider a downlink PA-assisted communication system adopting a NOMA scheme. Specifically, a base station (BS) feeds $N$ dielectric waveguides and a single PA is activated on each waveguide. Each waveguide is fed with different signals, which are sent to  \(K\) single-antenna users simultaneously. 
We adopt a three-dimensional (3D) Cartesian coordinate system to represent key elements in our system model. All \( N \) dielectric waveguides are elevated by distance \( d \) above the \( x \)-\( y \) plane and extend along the \( x \)-axis \cite{suzuki2022pinching,ding2024flexible}. Signals are fed from the BS to the beginnings of the waveguides.
 In the remainder of the paper, we refer to these starting points as feed points and their locations are denoted as $\bm{\psi}^{\text{FP}}_n = [0, D_n^{\text{FP}}, d]^{\text{T}}, \forall n\in \mathcal{N} = \{1,\cdots, N\}$. $D_n^{\text{FP}}$ denotes the location of the $n$-th feeding point along the $y$-axis. The positions of these $N$ PAs are represented as $\bm{\psi}^{\text{Pin}}_{n} = [x^{\text{Pin}}_{n}, D_n^{\text{FP}}, d]^{\text{T}},\forall n$, where $x^{\text{Pin}}_{n}$ denotes the location of the $n$-th PA along the $x$-axis. 
For the $K$ users, we assume that they are randomly distributed across the $x$-$y$ plane. The location of user\footnote{For simplicity, we assume that  users are static in this study. In our future work, we will optimize both the PAs' positions and beamforming based on the long-term user distribution.} $u_{k}, \forall k\!\in \!\mathcal{K} = \{1,\!\cdots\!,K\},$ is denoted as $\bm{\psi}_{k}^{\text{User}} = [x_{k}, y_{k}, 0]^{\text{T}}$, where $x_{k}$ and $y_{k}$ denote user $u_{k}$'s coordinates along the $x$-axis and $y$-axis, respectively.

\subsection{Signal Model}
Since multiple users are simultaneously served exploiting a NOMA-based approach, the transmitted signal $\bm{s}\in \mathbb{C}^{N\times 1}$ is a superposition of independent signals transmitted to $K$ users, which can be expressed as
\begin{align}\label{eq:s_a}
\bm{s} &= \sum\nolimits_{k=1}^{K}\bm{w}_k s_{k},
\end{align}where $\bm{w}_k\in \!\mathbb{C}^{N \times 1}, \bm{w}_k\! =\! [
               w_{k,1}, \cdots , w_{k,n} ,\cdots,w_{k,N}
             ]^{\text{T}}, \forall k,$ is the precoding vector for user $u_k$. We denote the maximum transmit power of the $n$-th dielectric waveguide as $P_{n}^{\max}$ such that $\sum_{k=1}^{K}|{w}_{k,n}|^2 \leq P_{n}^{\max},\forall n$.
 Also, $s_{k}\sim \mathcal{CN}(0,1)$ with $\mathbb{E}\{|s_{k}|^2\} = 1,\forall k,\mathbb{E}\{s_q^*s_k\}=0,\forall q\neq k,$  denotes the signal transmitted to user $u_{k}$.

\subsection{Channel Model}
As the PAs can be positioned in close proximity to users to establish near-field communications
\cite{ding2024flexible,suzuki2022pinching}, we adopt the spherical wave channel model \cite{9738442}. The end-to-end channel between the $n$-th feed point via its PAs and user $u_{k}$ is 
\begin{equation}\label{eq:h_uk_nopl}
\tilde{h}_{k,n} \!= \!\frac{\eta^{\frac{1}{2}}}{\|\bm{\psi}^{\text{User}}_{k} - \bm{\psi}^{\text{Pin}}_{n}\|} e^{-2\pi j \left(\frac{1}{\lambda}\|\bm{\psi}^{\text{User}}_{k} - \bm{\psi}^{\text{Pin}}_{n}\|+\frac{1}{\lambda_g}\|\bm{\psi}^{\text{FP}}_{n} - \bm{\psi}^{\text{Pin}}_{n}\|\right)}.
\end{equation}
Here, \(\eta = \frac{c^2}{16\pi^2 f_c^2}\), where \(c\) and \(f_c\) denote the speed of light and the signal carrier frequency, respectively. As all $N$ PAs are allocated along $N$ waveguides at a distance from their feed points, this inevitably imposes phase shifts on their emitted signals \cite{ding2024flexible}. To account for this phase-shifted behavior, we introduce a phase shift term, $e^{ \frac{-2\pi j}{\lambda_g}\|\bm{\psi}^{\text{FP}}_{n} - \bm{\psi}^{\text{Pin}}_{n}\|}$, to represent the phase shift incurred during the signal's propagation along the waveguide. The guided wavelength is defined as $\lambda_g = \frac{\lambda}{\eta_{\text{eff}}}$, where $\lambda = \frac{c}{f_c}$, and $\eta_{\text{eff}}> 1$ signifies the effective refractive index of the waveguide.
Besides, the phase shift $e^{\frac{-2\pi j }{\lambda}\|\bm{\psi}^{\text{User}}_{k} - \bm{\psi}^{\text{Pin}}_{n}\|}$ is due to the signal's wireless propagation from the $n$-th PA to user $u_k$.

Moreover, incorporating power attenuation along dielectric waveguides into the system model is essential for accurately evaluating and optimizing performance. 
Thus, the power attenuation for the $n$-th PA in lossy dielectric waveguide \cite{balanis2012advanced} is 
\begin{align}\label{eq:Pl}
\hspace{-2mm}P_{\text{Lossy},n}& = P_{n,0} e^{-2\alpha_{\text{D}}\,\|\bm{\psi}^{\text{FP}}_{n} - \bm{\psi}^{\text{Pin}}_{n}\|}, \forall n,\\
\alpha_{\text{D}}& = \lambda_g\varepsilon_{\text{r}}\pi f_c^2c^{-2}\tan \delta_e,\notag
\end{align} where  $\alpha_{\text{D}} \in \mathbb{R}^+,\varepsilon_{\text{r}}\in \mathbb{R}^+,$ and $\tan \delta_e\in \mathbb{R}^+$ are the attenuation constant, dielectric constant, effective electric loss tangent of the dielectric waveguide, respectively. $P_{n,0}$ is the initial power at the $n$-th feed point.
Combining the power attenuation \eqref{eq:Pl} in the dielectric waveguide, $\tilde{h}_{k,n}$ is now updated as
\begin{align}\label{eq:hn_uk_pl}
{h}_{k,n}& =(d_{k,n}^{\text{UPin}})^{-1}\eta^{\frac{1}{2}} e^{-\frac{2\pi j }{\lambda}d_{k,n}^{\text{UPin}}-(\frac{2\pi j }{\lambda_g}+\alpha_{\text{D}}) x^{\text{Pin}}_{n}},
\end{align}
 where $d_{k,n}^{\text{UPin}}= \sqrt{( x^{\text{Pin}}_n-x_k)^2+(D^{\text{FP}}_n-y_k)^2+d^2}$.
Hence, the equivalent channel from the PAs to user $u_k$, taking into account lossy dielectric waveguides, can be denoted as $\bm{h}_{k} = \begin{bmatrix}
                                                                                              {h}_{k,1} & \cdots & {h}_{k,n} &\cdots & {h}_{k,N}
                                                                                            \end{bmatrix}^{\text{T}}$.
\subsection{Received Signal at Users }
Combining \eqref{eq:s_a} and \eqref{eq:hn_uk_pl}, the received signal at  user \(u_{k}\), \(\forall k\), is given by
\begin{align}\label{eq:y_ba}
  y_{k} & =  \bm{h}_{k}^{\text{H}} \sum\nolimits_{m=1}^{K}\bm{w}_m s_{m} + n_{k}, \forall k,
\end{align} 
where $n_{k}\in \mathcal{CN}(0, \sigma_{k}^2), \forall k,$ is the thermal noise at user \(u_{k}\) and \(\sigma_{k}^2\) is the variance of \(n_{k}\). 

To implement NOMA in the considered PA-assisted system, we assume that users' channels are ordered in ascending strength based on their user index, without loss of generality \cite{7277111}. Specifically, the strongest user is denoted as \(u_{K}\) and the weakest user as \(u_{1}\), with all remaining users arranged in increasing order according to their indices. Consequently, as illustrated in Fig. \ref{fig:sys_mod}, user $u_k$ decodes the signals of all users $u_m, \forall m < k,$ and performs successive interference cancellation (SIC), while treating the signals of users $u_m, \forall m > k,$ as interference. User $u_K$ can mitigate interference from all other users by performing SIC, while the weakest user $u_1$ cannot decode any interference signals~\cite{7277111}.
According to NOMA principles, once user \(u_{k}\) has successfully eliminated the interference from users \(u_{m}\) with indices \(m < k, \forall m \in \{ 1, \cdots, k-1\} \) by applying the SIC operation, the achievable rate for \(u_{k}\) to decode its intended information is given by~\cite{7277111} 
\begin{align}\label{eq:R_kk}
R^{k}_{k} & = \log_2(1 + \text{SINR}^{k}_{k}), \quad \forall k.
\end{align} Here, \(\text{SINR}^{k}_{m}\) denotes the signal-to-interference-plus-noise ratio (SINR) at user \(u_{m}\) for decoding the message of user \(u_{k}\): 
\begin{align}\label{eq:SINR_bk}
\text{SINR}^{k}_{m}  = \frac{\abs{\bm{h}_m^{\text{H}}\bm{w}_k}^2}
{\sum_{b=k+1}^{K}\abs{\bm{h}_m^{\text{H}}\bm{w}_b}^2+\sigma_m^2}, \forall k,m. 
\end{align} It is worth mentioning that to ensure that the data rate in \eqref{eq:R_kk} achieves a target data rate \(R_{\text{th}}\) for user \(u_{k}\) (i.e., $R^{k}_{k}\geq R_{\text{th}}$), the data rate of all users \(u_{m}, \forall m > k\), decoding the message of user \(u_{k}\), denoted as \(R^{k}_{m}\), must be at least \(R_{\text{th}}\) for \(u_{k}\), i.e.,
\begin{align}\label{eq:R_kb} 
R^{k}_{m}\!= \!\log_2(1\! + \!\text{SINR}^{k}_{m}) \!\geq\! R_{\text{th}}, m \in\{ k+1,\cdots,K\}, \forall k.
\end{align}
To further elaborate, \eqref{eq:R_kb} ensures that the SINR for user $u_m$ when decoding the message intended for user $u_k$ (i.e., $\text{SINR}_m^k$ corresponding to $R_m^k$, where $m > k$) exceeds the SINR for user $u_k$ decoding its own message (i.e., $\text{SINR}_k^k$ corresponding to $R_k^k$). Once this condition is satisfied, users with stronger channels (i.e., those with higher indices in the predefined ordering) can successfully perform SIC. Based on this observation, we propose maximizing the minimum SINR among $\text{SINR}_k^k$ and $\{\text{SINR}_m^k, \forall m > k\}$. Then, the users' sum-rate is 
\begin{align}\label{eq:R_sum}
 & R_{\text{sum}} = \sum\nolimits_{k=1}^{K}R_{k}, \\[1mm]
&R_{k}=\begin{cases} \!\!
 \log_2( 1\! \!+\! \min \{ \!\text{SINR}_{k}^{k},\text{SINR}_{k+1}^{k}, \!\cdots\!, \text{SINR}_{K}^{k} \!\}), \!1\!\leq\! k \!<\!K, \\
\!\!\log_2 ( 1 + \frac{\abs{\bm{h}_k^{\text{H}}\bm{w}_k}^2}{\sigma_{K}^2}) ,  k =K.
\end{cases}  \notag
\end{align}
It can be observed from \eqref{eq:R_sum} that the effective channel strength for users can be controlled by adjusting the positions of the \(N\) PAs, $x^{\text{Pin}}_{n}$, and by altering the precoding vector to the PAs, \(\bm{w}_k\). Consequently, the following condition must be satisfied \cite{7277111}:
\begin{align}\label{eq:channelorders}
\abs{\bm{h}_k^{\text{H}}\bm{w}_1}^2 
\geq& \cdots 
\geq \abs{\bm{h}_k^{\text{H}}\bm{w}_m}^2  
\geq \cdots 
\geq \abs{\bm{h}_k^{\text{H}}\bm{w}_K}^2, \forall k.
\end{align}
Specifically, \eqref{eq:channelorders} ensures that users with stronger channels receive a lower combined channel gain and reduced beamforming power, thereby optimizing the SINRs necessary for decoding the messages of other users. Thus, \eqref{eq:channelorders} plays a critical role in achieving efficient resource allocation and enhancing overall system performance.

\section{Problem Formulation}
In this section, we formulate a problem to maximize the sum-rate of the proposed PA-assisted NOMA system. Our problem formulation takes into account the power attenuation over lossy dielectric waveguides, which is expressed as \eqref{proposed_formulation_origion} by optimizing the locations for $N$ PAs, $x^{\text{Pin}}_{n}$, and the precoding vectors, \(\bm{w}_k\):
\begin{align}\label{proposed_formulation_origion}
  \underset{x^{\text{Pin}}_{n}, \bm{w}_k}{\mathrm{maximize}} \,\,&  R_{\text{sum}}\\
  \mathrm{s.t.}\,\,
    &\mathrm{C1}\hspace{-1mm}: \eqref{eq:channelorders},\notag\\[-0mm]
 & \mathrm{C2}\hspace{-1mm}: \sum\nolimits_{k=1}^{K}\abs{{w}_{k,n}}^2 \leq P_{n}^{\max} ,\forall n,
\notag\\[-0mm]
 &\mathrm{C3}\hspace{-1mm}: R_{k}\geq R_{\min}, \forall k,\notag\\[-0mm]
 & \mathrm{C4}\hspace{-1mm}: 0\leq x^{\text{Pin}}_{n} \leq x^{\max},\forall n.
\notag
\end{align}
Specifically, to facilitate the decoding process, constraint $\mathrm{C1}$ ensures that the channels are ordered by strength.
Constraint $\mathrm{C2}$ guarantees that the transmit power consumption of the $n$-th dielectric waveguide does not exceed its maximum power budget, $P_{n}^{\max} $.  Constraint $\mathrm{C3}$ restricts that the achievable rate of user $u_k$ is not less than the minimum rate requirement $R_{\min}$. Constraint $\mathrm{C4}$ restricts the movement of the PAs along the $x$-axis to the range $[0, x^{\max}]$.
The formulated problem is non-convex due
to the coupling between optimization variables $\bm{w}_k$ and $x_{n}^{\text{pin}}$ in the objective function, constraints $\mathrm{C1}$, and $\mathrm{C3}$.  
In general, obtaining the globally optimal solution to \eqref{proposed_formulation_origion} necessitates the use of a brute-force search, which becomes computationally infeasible even for systems of a moderate size. As a practical alternative, we propose a computationally efficient suboptimal iterative algorithm based on AO in the following section.

\section{Solution Of The Optimization Problem}

\subsection{Problem Transformation}
Firstly, we introduce constraints $\mathrm{C5}$, $\mathrm{C6}$, and slack optimization variables, $r_k,\forall k$, to replace $1 + \min \{ \text{SINR}_{k}^{k}, \cdots, \text{SINR}_{K}^{k} \}$, $k \in\mathcal{K}\setminus\{ K\}$, and $1 + \abs{\bm{h}_K^{\text{H}}\bm{w}_K}^2\sigma_{K}^{-2}$  in $R_{\text{sum}}$ and $R_k$. As such, constraints $\mathrm{C5}$ and $\mathrm{C6}$ are equivalently given by 
\begin{align}\label{eq:rk}
&\mathrm{C5}\hspace{-1mm}:r_k -1 \leq \min\{  \text{SINR}_{k}^{k}, \cdots, \text{SINR}_{K}^{k}  \}, k \in\mathcal{K}\!\setminus\!\{ K\},
\\
& \Leftrightarrow r_k\! -\!1 \!\leq  \text{SINR}_{m}^{k}, m\!\in\!\{ k, \cdots,\! K-1\},k \in\mathcal{K}\!\setminus\!\{ K\},\notag\\
&\mathrm{C6}\hspace{-1mm}: r_K -1 \leq \sigma_{K}^{-2}\hspace{1mm}\text{Tr}(\bm{h}_K^{\text{H}}\bm{w}_K\bm{w}_K^{\text{H}}\bm{h}_K),\notag 
\end{align} 
respectively. To handle the non-convexity, we introduce slack variables $\xi_{m,k}$ to decouple the fractional form in SINRs in \eqref{eq:rk}. Then, $\mathrm{C5}$  can be equivalently transformed as 
\begin{align}
    &{\mathrm{C5a}} \hspace{-1mm}: \xi_{m,k} r_k - \xi_{m,k} \leq \text{Tr}(\bm{h}_{m}^{\text{H}} \bm{w}_k\bm{w}_k^{\text{H}}\bm{h}_{m}),\\
    & \hspace{7mm} m\in\{ k, \cdots, K-1\}, k \in\mathcal{K}\!\setminus\!\{ K\},\notag \\
    &{\mathrm{C5b}} \hspace{-1mm}: \sum\nolimits_{b=k+1}^{K} \text{Tr}(\bm{h}_{m}^{\text{H}} \bm{w}_b\bm{w}_b^{\text{H}} \bm{h}_{m}) + \sigma_m^2 \leq \xi_{m,k}, \notag\\
    & \hspace{7mm} m \in\{ k, \cdots, K-1\}, k \in\mathcal{K}\!\setminus\!\{ K\}.\notag
\end{align}
On the other hand, constraint $\mathrm{C1},  \forall k,m = \{2,\cdots, K\},$ can be equivalently transformed as following: 
\begin{align}
\hspace{-3mm}\mathrm{C1} 
\Leftrightarrow \overline{\mathrm{C1}}:  \text{Tr}(\bm{h}_k^{\text{H}}\bm{w}_{m}\bm{w}_{m}^{\text{H}}\bm{h}_k) 
\leq  \text{Tr}(\bm{h}_k^{\text{H}}\bm{w}_{m-1}\bm{w}_{m-1}^{\text{H}}\bm{h}_k).
\end{align}
By defining $\bm{W}_k \stackrel{\triangle}{=} \bm{w}_k\bm{w}_k^{H}$, the optimization problem \eqref{proposed_formulation_origion} can be equivalently rewritten as 
\begin{align}\label{eq:pf_v1}
  &\underset{x^{\text{Pin}}_{n}, \bm{W}_k\in\mathbb{H}^{N},\xi_{m,k}, r_k}{\mathrm{maximize}} \,\, 
\sum\nolimits_{k=1}^{K}\log_2(r_k )\\
  \mathrm{s.t.}\,\,
      & \overline{\mathrm{C1}}, \notag\\[-0mm]
    &\overline{\mathrm{C2}}\hspace{-1mm}: \sum\nolimits_{k=1}^{K}\!\!\!\bm{W}_{k}[n,n] \leq P_{n}^{\max} ,\forall n,
    \notag\\[-0mm]
    & \overline{\mathrm{C3}}\hspace{-1mm}: r_k  \geq 2^{R_{\min}}, \forall k, \notag\\[-0mm]
    &{\mathrm{C4}},{\mathrm{C5a}},{\mathrm{C5b}}, \notag\\[-0mm]
    &\overline{\mathrm{C6}}\hspace{-1mm}: r_K -1 \leq \text{Tr}(\bm{h}_K\bm{h}_K^{\text{H}}\bm{W}_K)/\sigma_{K}^2,\notag\\[-0mm]
   &\mathrm{C7}\hspace{-1mm}: \text{Rank}(\bm{W}_k)\leq 1,\forall k, \notag\\[-0mm]
    &\mathrm{C8}\hspace{-1mm}: \bm{W}_k \succeq \bm{0},\forall k.\notag
\end{align}

Note that in the optimization problem in \eqref{eq:pf_v1}, constraints $\overline{\mathrm{C1}}$, $\overline{\mathrm{C5a}}$, $\overline{\mathrm{C5b}}$, and $\overline{\mathrm{C6}}$ are nonconvex due to variable couplings, while constraint $\mathrm{C7}$ is a discrete rank constraint. Hence, in the following section, an iterative AO algorithm is introduced to obtain a suboptimal solution to \eqref{eq:pf_v1}. Specifically, the proposed algorithm addresses the coupled variables $\bm{W}_k$ and $x^{\text{Pin}}_{n}$ by decomposing \eqref{eq:pf_v1} into two subproblems. Then, the proposed algorithm alternately updates $\bm{W}_k$ or $x^{\text{Pin}}_{n}$ while fixing the remaining variables in each sub-problem.

\subsection{Sub-problem 1: Optimization of the Precoding Vector} \label{sec:subp1}
\begin{table}[t]
\vspace{-4mm}
\scriptsize
\linespread{0.9}
\begin{algorithm} [H]
\caption{\small  SCA-based Iterative Precoder Optimization} \label{alg_1}
\begin{algorithmic} [1]
\STATE Set the maximum iterations number $t_{\max}$,  initial the index of the previous iteration $t_1=0$,   and optimization variables in $\xi_{m,k}^{(t_1)},\forall k,m,$ and $r_k^{(t_1)},\forall k$, for a given constant $\bm{h}_{k}^{\mathrm{Con.}}$.
\REPEAT[Main Loop: SCA]
\STATE Solve  \eqref{eq:sp1_v1} with given optimization variables in $\xi_{m,k}^{(t_1)}$  and $r_k^{(t_1)}$ and constants $\bm{h}_k^{\mathrm{Con.}}$ to obtain the variables for  $\xi_{m,k}^{(t_1+1)}$ and $r_k^{(t_1+1)}$;
\STATE Set $t_1=t_1+1$ and update $\xi_{m,k}^{(t_1)}$ and $r_k^{(t_1)}$;
\UNTIL{convergence or $t_1=t_{\max}$}.
\end{algorithmic}
\end{algorithm}\vspace{-7mm}
\end{table}
In this section, we focus on optimizing the precoding matrix $\bm{W}_k$ and slack variables $\xi_{m,k}$ and $r_k$, assuming a fixed and feasible location for the PAs. With this assumption, the channel $\bm{h}_k$ remains constant and sub-problem 1 is given by 
\begin{align}\label{eq:pf_ori}
  \underset{\bm{W}_k\in\mathbb{H}^{N},\xi_{m,k}, r_k}{\mathrm{maximize}} \,\,& 
\sum\nolimits_{k=1}^{K}\log_2(r_k )
   \\
  \mathrm{s.t.}\,\,
      & \overline{\mathrm{C1}},\overline{\mathrm{C2}},\overline{\mathrm{C3}},{\mathrm{C5a}}, {\mathrm{C5b}},\overline{\mathrm{C6}},\mathrm{C7},\mathrm{C8}.\notag
\end{align}
Consequently, only constraints ${\mathrm{C5a}}$ and $\mathrm{C7}$ remain non-convex in this sub-problem.
Firstly, we apply an iterative method based on successive convex approximation (SCA) to handle the difference of convex (d.c.) functions in ${\mathrm{C5a}}$. For any feasible point $\xi_{m,k}^{(t_1)}$ and $r_k^{(t_1)}$, where $(t_1)$ denotes the iteration index of SCA, as summarized in \textbf{Algorithm \ref{alg_1}}, we establish an upper bound function for $-\xi_{m,k}^2,m \in\{ k, \!\cdots\!, K\!-1\},$ and $-r_k^2,k \in\mathcal{K}\setminus\!\{K\},$ by adopting their first-order Taylor series expansions:
\begin{align} \label{eq:taylor_xiR}
-0.5\xi_{m,k}^2 \leq& -0.5(\xi_{m,k}^{(t_1)})^2 - \xi_{m,k}^{(t_1)}(\xi_{m,k}-\xi_{m,k}^{(t_1)}), \\[-0mm]
-0.5r_k^2 \leq& -0.5(r_k^{(t_1)})^2 - r_k^{(t_1)}(r_k-r_k^{(t_1)}). \notag
\end{align}
As such, a convex subset of ${{\mathrm{C5a}}}$ can be derived as 
\begin{align}
   {\overline{\mathrm{C5a}}}\hspace{-1mm}: & -\text{Tr}(\bm{h}_m\bm{h}_m^{\text{H}}\bm{W}_{k})+ 0.5(\xi_{m,k}+ r_k)^2\\[1mm]
  &- \xi_{m,k}^{(t_1)}(\xi_{m,k}-\xi_{m,k}^{(t_1)})-0.5(r_k^{(t_1)})^2 - r_k^{(t_1)}(r_k-r_k^{(t_1)}) \notag\\[1mm]
  &- \xi_{m,k} -0.5(\xi_{m,k}^{(t_1)})^2 \leq 0,  m \in\{ k, \cdots, K-1\}, \forall k,\notag
\end{align}
such that $\overline{\mathrm{C5a}}$ implies ${{\mathrm{C5a}}}$.
Currently, rank constraint $\mathrm{C7}$ is the only non-convex component of sub-problem 1. To tackle this, we apply the semidefinite relaxation (SDR) technique \cite{hu2021robust}, removing the rank constraint. Thus, we have 
\begin{align}\label{eq:sp1_v1}
  \underset{\bm{W}_k\in\mathbb{H}^{N},\xi_{m,k}, r_k}{\mathrm{maximize}} \,\,& 
\sum\nolimits_{k=1}^{K}\log_2(r_k )
   \\[-0mm]
  \mathrm{s.t.}\,\,
      & \overline{\mathrm{C1}},\mathrm{C2},\overline{\mathrm{C3}},{\overline{\mathrm{C5a}} },{\mathrm{C5b}} ,\overline{\mathrm{C6}},\mathrm{C8}.\notag
\end{align}
At this stage, the problem in \eqref{eq:sp1_v1} has been transformed into a convex semidefinite program, which can be efficiently solved exploiting standard convex optimization numerical solvers, e.g., CVX \cite{grant2014cvx}. The tightness of the applied SDR is analyzed in the following theorem.

\begin{theorem}
For $P^{\max}_n>0,\forall n,$ and if \eqref{eq:sp1_v1} is feasible, a rank-one solution of \eqref{eq:sp1_v1} can always be constructed.
\end{theorem}

\emph{\quad Proof: }
Due to space limitation, we provide only a brief sketch of the proof. By examining the Karush-Kuhn-Tucker (KKT) conditions of \eqref{eq:sp1_v1}, it can be demonstrated that a rank-one solution $\mathbf{W}_k$ must exist to ensure a bounded solution to the dual problem of \eqref{eq:sp1_v1}. Furthermore, a rank-one solution to \eqref{eq:sp1_v1} can be explicitly constructed by leveraging the dual variables of its corresponding dual problem.
 \qed

By applying the SCA technique, solving \eqref{eq:sp1_v1} yields a suboptimal solution to \eqref{eq:pf_ori}. We further refine this by iteratively updating the feasible solution by solving  \eqref{eq:sp1_v1}  at each \( t_1 \)-th iteration.
 The proposed SCA-based \textbf{Algorithm \ref{alg_1}} converges to a suboptimal solution, with proof in \cite{opial1967weak}.

\subsection{Sub-problem 2: PAs' Location Optimization}

\begin{table}[t]
\vspace{-4mm}
\scriptsize
\linespread{0.9}
\begin{algorithm} [H]
\caption{\small  SCA-based Iterative PAs' Location Optimization} \label{alg_2}
\begin{algorithmic} [1]
\STATE Set the maximum iterations number $t_{\max}$, initial the index of the previous iteration $t_2=0$, and optimization variables in  $x^{\text{Pin}^{(t_2)}}_{n},\gamma_{k,i,q}^{(t_2)}, r_k^{(t_2)}$, and $\xi_{m,k}^{(t_2)},$ for a given $\mathbf{W}_k^{\mathrm{Con.}}$.
\REPEAT[Main Loop: SCA]
\STATE Solve \eqref{eq:sp2_v1} with given optimization variables in $x^{\text{Pin}^{(t_2)}}_{n},  \gamma_{k,i,q}^{(t_2)},r_k^{(t_2)}$, and $\xi_{m,k}^{(t_2)}$ and constants $\mathbf{W}_k^{\mathrm{Con.}}$ to obtain the variables for $x^{\text{Pin}^{(t_2+1)}}_{n}, \gamma_{k,i,q}^{(t_2+1)}, r_k^{(t_2+1)}$, and $\xi_{m,k}^{(t_2+1)}$;
\STATE \hspace{0mm}Set $t_2\!=\!t_2\!+\!1$ and update $x^{\text{Pin}^{(t_2)}}_{n}\hspace{-2mm}, \gamma_{k,i,q}^{(t_2)}, r_k^{(t_2)}$, and $\xi_{m,k}^{(t_2)}$;
\UNTIL{convergence or $t_2=t_{\max}$}.
\end{algorithmic}
\end{algorithm}
\vspace{-7mm}
\end{table}
Now, we fix the precoding matrix $\bm{W}_k, \forall k$, and optimize the location of the PAs, i.e., $x^{\text{Pin}}_{n},\forall n$, slack variables $\xi_{m,k},\forall m,k$, and $r_k,\forall k$, to improve the channel condition $\bm{h}_k, \forall k$. 
Therefore, sub-problem 2 is written as 
\begin{align}\label{eq:sp2_ori}
  \underset{x^{\text{Pin}}_{n}, \xi_{m,k}, r_k}{\mathrm{maximize}} \,\,& 
\sum\nolimits_{k=1}^{K}\log_2(r_k )
   \\
  \mathrm{s.t.}\,\,
      & \overline{\mathrm{C1}},\overline{\mathrm{C3}},\mathrm{C4},{\mathrm{C5a}}, {\mathrm{C5b}},\overline{\mathrm{C6}}.\notag
\end{align}
To address the non-convexity of \eqref{eq:sp2_ori}, $\text{Tr}(\bm{h}_k\bm{h}_k^{\text{H}}\bm{W}_m)$ is equivalently rewritten as $\sum\nolimits_{i=1}^{N}\sum\nolimits_{q=1}^{N} h_{k,i}$$ h_{k,q}^* \bm{W}_m[q,i]$.
Since $\text{Tr}(\bm{h}_k\bm{h}_k^{\text{H}}\bm{W}_m)$ is a real number, thus we have
\begin{align}
&\text{Tr}(\bm{h}_k\bm{h}_k^{\text{H}}\bm{W}_m) = \sum\nolimits_{i=1}^{N}\sum\nolimits_{q=1}^{N} \mathcal{R}\{h_{k,i} h_{k,q}^*\} \bm{W}_m[q,i] \\
&= \sum\nolimits_{i=1}^{N}\sum\nolimits_{q=1}^{N} \frac{d_{k,i}^{\text{UPin}}d_{k,q}^{\text{UPin}}}{\eta e^{-\alpha_{\text{D}}(x^{\text{Pin}}_i+x^{\text{Pin}}_q)}  \bm{W}_m[q,i]}, \forall k,m.\notag
\end{align}
Now, we introduce slack optimization variables $\tau_{k,i,q}$ and $\gamma_{k,i,q}$, $\forall k, i,q$, and constraints  $\mathrm{C9}$ and $\mathrm{C10}$, which are 
\begin{align}
         & {\mathrm{C9}}\hspace{-1mm}:
        -\alpha_{\text{D}}x^{\text{Pin}}_i-\ln(d_{k,q}^{\text{UPin}})-\alpha_{\text{D}}x^{\text{Pin}}_q -\ln(d_{k,i}^{\text{UPin}}) \hspace{3mm} \notag\\[1mm]
        &\hspace{6mm} -\ln( \tau_{k,i,q})+\ln(\eta)\leq 0, \forall k,i,q, \text{  and }\notag\\[1mm]
        & {\mathrm{C10}}\hspace{-1mm}: 
\alpha_{\text{D}}x^{\text{Pin}}_i\!+\!\ln(d_{k,i}^{\text{UPin}}) +\alpha_{\text{D}}x^{\text{Pin}}_q\!+\!\ln(  d_{k,q}^{\text{UPin}} )\!   \notag\\[1mm] &\hspace{6mm} +\!\ln( \gamma_{k,i,q})\!-\!\ln(\eta)\! \leq\!  0, \forall k,i,q,  
\end{align}
respectively. By following the same approach as for handling sub-problem 1 in Section \ref{sec:subp1}, we apply the SCA to address the non-convexity in constraints $\overline{\mathrm{C1}}, {\mathrm{C5a}},{\mathrm{C5b}},\overline{\mathrm{C6}}$, $\mathrm{C9}$, and $\mathrm{C10}$. Thus, the optimization problem in \eqref{eq:sp2_ori} can be transformed as the optimization problem in \eqref{eq:sp2_v1} to obtain a suboptimal solution: 
\begin{align}\label{eq:sp2_v1}
    &\underset{x^{\text{Pin}}_{n}, \tau_{k,i,q},\gamma_{k,i,q},\xi_{m,k}, r_k}{\mathrm{maximize}} \,\, 
    \sum\nolimits_{k=1}^{K}\log_2(r_k )\\
    \mathrm{s.t.}\,\,
    & \widetilde{\overline{\mathrm{C1}}}\hspace{-1mm}:  \sum_{i=1}^{N} \sum_{q=1}^{N}\tau_{k,i,q}\bm{W}_m[q,i] -\sum_{i=1}^{N} \sum_{q=1}^{N}\gamma_{k,i,q}\bm{W}_{m-1}[q,i] 
    \notag\\
    &\hspace{6mm}\leq  0,  m \in\{2,\!\cdots\!,K\},\forall k, \overline{\mathrm{C3}},\mathrm{C4}, \notag \\
    &\widetilde{{\mathrm{C5a}}} \hspace{-1mm}: 0.5(\xi_{m,k}+ r_k)^2-0.5(\xi_{m,k}^{(t_2)})^2 -0.5(r_k^{(t_2)})^2 \notag\\[1mm]
    &\hspace{6mm} - \xi_{m,k}^{(t_2)}(\xi_{m,k}-\xi_{m,k}^{(t_2)})- r_k^{(t_2)}(r_k-r_k^{(t_2)}) \notag\\
    &\hspace{6mm}-\sum_{i=1}^{N} \sum_{q=1}^{N}\gamma_{m,i,q}\bm{W}_{k}[q,i]- \xi_{m,k}\leq 0, \notag\\
    &\hspace{6mm} m \in\{ k, \cdots, K-1\},k\in\mathcal{K}\setminus\{K\},\notag \\
    & \widetilde{{\mathrm{C5b}}} \hspace{-1mm}: \sum_{b=k+1}^{K}  \sum_{i=1}^{N} \sum_{q=1}^{N}\tau_{m,i,q}\bm{W}_b[q,i] + \sigma_m^2 \leq \xi_{m,k},\notag\\
    &\hspace{6mm} m \in\{k, \cdots, K-1\},k\in\mathcal{K}\setminus\{K\},\notag\\
     &\widetilde{\overline{\mathrm{C6}}}\hspace{-1mm}:\sigma_{K}^2 r_K -\sigma_{K}^2\leq \sum\nolimits_{i=1}^{N} \sum\nolimits_{q=1}^{N}\gamma_{K,i,q}\bm{W}_K[q,i],\notag\\
         & {\mathrm{C9}}\hspace{-1mm}:
        -\alpha_{\text{D}}x^{\text{Pin}}_i+f_{\text{sca}}(-\ln
        (d_{k,q}^{\text{UPin}}
        ))-\alpha_{\text{D}}x^{\text{Pin}}_q \notag\\[1mm]
   &\hspace{6mm}+f_{\text{sca}}(-\ln
        (d_{k,i}^{\text{UPin}}
        ))+\ln(\eta)\leq \ln
        ( \tau_{k,i,q}
        ), \forall k,i,q,\notag\\[1mm]
        & {\mathrm{C10}}\hspace{-1mm}: 
   \!\alpha_{\text{D}}x^{\text{Pin}}_i\!+\!\ln
   (d_{k,i}^{\text{UPin}}
   )\!+\!\alpha_{\text{D}}x^{\text{Pin}}_q\!+\!\ln(  d_{k,q}^{\text{UPin}} )\notag\\[1mm]
   &\hspace{6mm} +\!f_{\text{sca}}(\ln( \gamma_{k,i,q}))\!-\!\ln(\eta)\! \leq\!  0, \forall k,i,q,\notag 
\end{align}
where 
\begin{align}
&f_{\text{sca}}(-d_{k,i}^{\text{UPin}})
\!  = \! -d_{k,i}^{\text{UPin}} |_{x_n^{\text{Pin}^{(t2)}}}\!\! - \frac{x^{{\text{Pin}^{(t_2)}}}_i - x_k}{ d_{k,i}^{\text{UPin}} |_{x_n^{\text{Pin}^{(t2)}}}} (x^{\text{Pin}}_i\! -\! x_i^{\text{Pin}^{(t_2)}}\!),\notag\\
 & f_{\text{sca}}(\ln( \gamma_{k,i,q} )) = \ln( \gamma_{k,i,q} ^{(t_2)} ) + \frac{1}{ \gamma_{k,i,q} ^{(t_2)}} ( \gamma_{k,i,q} - \gamma_{k,i,q} ^{(t_2)})\text{, and} \notag\\
 & f_{\text{sca}}(-\ln( d_{k,q}^{\text{UPin}} )) =-\ln( d_{k,i}^{\text{UPin}} |_{x_n^{\text{Pin}^{(t2)}}}) 
\notag\\&\hspace{28mm}- \frac{x_q^{\text{pin}^{(t_2)}} - x_k}{(d_{k,i}^{\text{UPin}} |_{x_n^{\text{Pin}^{(t2)}}})^2} (x^{\text{Pin}}_q - x_q^{\text{pin}^{(t_2)}}).\notag\end{align}
Now, the optimization problem in \eqref{eq:sp2_v1} can be efficiently solved adopting a standard convex programming solver, yielding a suboptimal solution to the original problem in \eqref{eq:sp2_ori}.
The proposed algorithm for solving \eqref{eq:sp2_v1} is presented in \textbf{Algorithm \ref{alg_2}}. Furthermore, the overall process, which iteratively solves the two sub-problems in \eqref{eq:sp1_v1} and \eqref{eq:sp2_v1}, is summarized in \textbf{Algorithm \ref{alg_overall}}. 
The proposed \textbf{Algorithm \ref{alg_overall}} is guaranteed to converge to a suboptimal solution of \eqref{proposed_formulation_origion} within a polynomial-time computational complexity \cite{polik2010interior}.

\begin{table}[t]
\vspace{-3mm}
\scriptsize
\linespread{0.9}
\begin{algorithm} [H]
\caption{\small Overall AO Algorithm} \label{alg_overall}
\begin{algorithmic} [1]
\STATE Set the maximum iterations number $\tau_{\max}$, initial the index of the previous iteration $\tau=0$, and the optimization variables in $\mathbf{W}_k^{(\tau)}$ and $x^{\text{Pin}^{(\tau)}}_{n}$.
\REPEAT[Main Loop: AO]
\STATE Obtain $\mathbf{W}_k^{(\tau+1)}$ by \textbf{Algorithm \ref{alg_1}}  with given optimization variables in constants $\bm{h}_{k}^{\mathrm{Con.}} =\bm{h}_{k}|_{x^{\text{Pin}}_{n}=x^{\text{Pin}^{(\tau)}}_{n}}$.
\STATE Obtain the variables  $x^{\text{Pin}^{(\tau+1)}}_{n}$ by \textbf{Algorithm \ref{alg_2}} with given $\mathbf{W}_k^{(\tau+1)}$;
\STATE Update $x^{\text{Pin}^{(\tau)}}_{n}$.
\STATE Set $\tau=\tau+1$ and update the optimization variables;
\UNTIL{convergence or $\tau=\tau_{\max}$}.
\end{algorithmic}
\end{algorithm}
\vspace{-7mm}
\end{table}

\section{Numerical Results}

\begin{figure}[t]
  \centering 
  \includegraphics[width=0.43\textwidth]{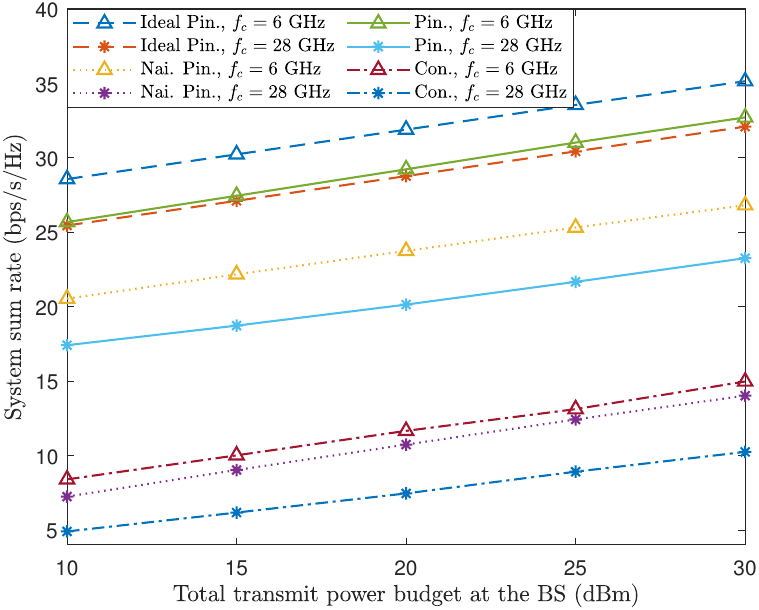}
    \vspace{-0mm}
  \caption{System sum-rate versus the total transmit power budget of pinching or conventional antennas at the BS, $P_{\max}$, with $K=6$ and $N=2$.}
  \label{fig:diffP_sumR}
\vspace{-0mm}
\end{figure}

\begin{figure}[t]
  \centering
  \includegraphics[width=0.43\textwidth]{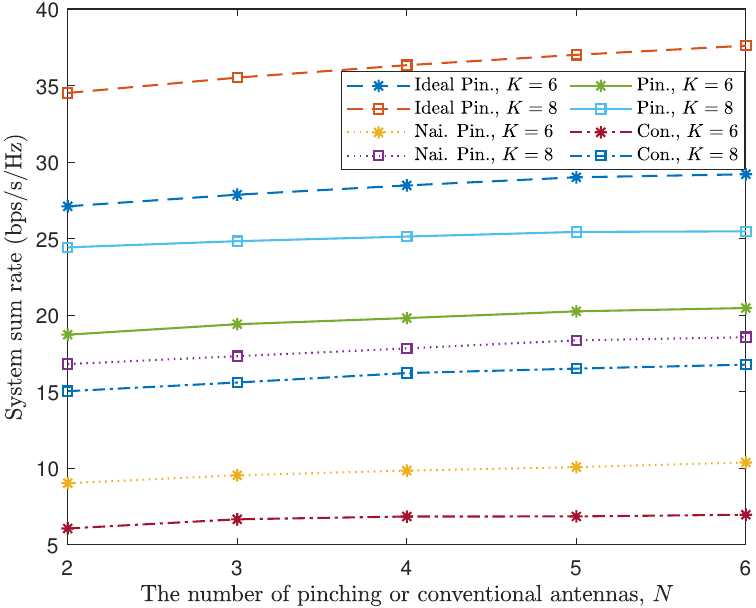}
\vspace{-0mm}
  \caption{System sum-rate versus the number of pinching or conventional antennas at the BS, $N$, with $P_{\max} = 15$ dBm and $f_c = 28$ GHz.}
  \label{fig:diffN_sumR}
\vspace{-0mm}
\end{figure}

This section evaluates the system performance of the proposed PA-assisted NOMA system with multiple dielectric waveguides via simulation based on the setup in Fig. \ref{fig:sys_mod}. Specifically, each dielectric waveguide is spaced 10 meters apart, i.e., $\left| D_{n+1}^{\text{FP}} - D_n^{\text{FP}} \right| = 10$ \text{meters}, $n \in \{ 1, \dots, N-1\}$,
and \( d = 3 \) meters \cite{ding2024flexible}.  $x^{\max}$ is set to $100$ meters. The minimum rate requirement $R_{\min}$ for users is $0.5$ bps/s/Hz. The $K$ users are randomly distributed within a square area of side length 100 m. Without loss of generality, the maximum power budget of each dielectric waveguide, $P^{\max}_n$, is assumed to be identical and the total transmitted power is given by  
$P_{\max} = \sum_{n=1}^{N} P^{\max}_n.$
The dielectric waveguides are composed of Polytetrafluoroethylene (PTFE)\cite{suzuki2022pinching}, with key parameters specified as follows: $\eta_{\text{eff}} = 1.42$, $\varepsilon_r=2.1$, and $\tan \delta_e = 2\times10^{-4}$. The values of \( P_{\max} \), \( K \), \( N \), and the carrier frequency \( f_c \) are specified in each figure.
For comparison, in addition to the proposed scheme, denoted as “Pin.” in Figs. \ref{fig:diffP_sumR} and \ref{fig:diffN_sumR}, we also evaluate the system performance of three other schemes: 
1) Ideal Pin. scheme: This scheme is identical to the proposed scheme but assumes the waveguides are deployed with perfect dielectric material, thereby ensuring no power attenuation through the waveguides. Thus, it serves as a performance upper bound;
2) Nai. Pin. scheme: This is a naive baseline scheme, which assumes that there is no power attenuation through the waveguides during the resource allocation process. To ensure a fair comparison, the system sum-rate was recalculated based on the channel model considering power attenuation in \eqref{eq:hn_uk_pl}. Furthermore, we focus on its upper-bound performance by assuming that this scheme is not required to satisfy the channel ordering constraint $\mathrm{C1}$, thereby exploiting the feasible solution set to facilitate performance evaluation;
3) Con. scheme: This is a baseline scheme adopting a conventional antenna system, where \( N \) BS antennas are fixed at the center of the feed points. A free-space channel model is adopted between the BS and each user under far-field conditions.

Fig. \ref{fig:diffP_sumR} illustrates the system sum-rate versus the total transmit power budget for both pinching and conventional antennas at the BS across different schemes.  
As expected, the sum-rate for all schemes increases monotonically with \( P_{\max} \). Moreover, the proposed scheme outperforms the conventional antenna scheme. This performance gain stems from  the ability of PAs to establish a more favorable radio propagation environment by optimizing their pinching positions. Consequently, the large-scale path loss between the BS and users can be effectively reduced compared with a system adopting conventional antennas.  
On the other hand, the proposed scheme exhibits a performance gap compared to the ideal PA scheme. This discrepancy arises because the ideal PA scheme assumes perfect dielectric waveguides with no power loss during signal propagation.
However, power attenuation is unavoidable in reality. Indeed, the naive PA scheme neglects power attenuation, causing a mismatch between the optimized beamformer and PA positions with the actual channel state information, which leads to performance degradation compared to the proposed scheme.
Notably, at high carrier frequencies \( f_c \), the performance degradation becomes more significant, particularly at mmWave frequencies such as 28 GHz. Indeed, the \( f_c^2 \)-dependent term in the attenuation constant \( \alpha_{\text{D}} \) in \eqref{eq:Pl} leads to exponential signal decay as frequency increases, resulting in more severe power attenuation over the dielectric waveguide. This exacerbates the discrepancy between the actual channel \( h \) and the one assumed in the naive scheme, causing the PAs' positions to deviate from the optimized placement and further degrading system performance.

Fig. \ref{fig:diffN_sumR} illustrates the system sum-rate  versus the number of pinching or conventional antennas at the BS, across different schemes. As \( N \) increases, the system sum-rate improves. This is because a larger \( N \) provides more DoF for optimizing the channel between the PAs (or conventional antennas) and the users, thereby enhancing the beamforming gain. However, the performance improvement gradually saturates for a sufficiently large $N$, as the performance is ultimately constrained by the power budget. We also observe that increasing the number of users, $K$, enables better spectrum utilization in the NOMA system, thereby enhancing the system sum-rate.


\section{Conclusion}
This paper investigated a PA-assisted multiuser MISO NOMA wireless system employing multiple dielectric waveguides,  taking into account the crucial factor of power attenuation over dielectric waveguides. By jointly optimizing the precoding vector and the placement of the PAs, we formulated a non-convex optimization problem for maximizing the system's sum-rate. An AO-based algorithm was proposed, yielding an effective suboptimal solution. Our results reveal significant performance gains achieved by the proposed PA-assisted system compared to both conventional antennas and a naive PA-assisted system that ignores power attenuation in lossy dielectric waveguides.

\bibliographystyle{IEEEtran}
\bibliography{pinching_ant_multiWaveguide_NOMA}

\end{document}